# Blind separation of rotor vibration signals in high-noise environments

Pengfei Xu[a,*], Yinjie Jia[a,b] and Zhijian Wang[a]

[a]College of Computer and Information, Hohai University, Nanjing, Jiangsu 210098, China
[b]Faculty of Electronic Information Engineering, Huaiyin Institute of Technology, Huaian, Jiangsu 223003, China

**Abstract:** During the operation of the engine rotor, the vibration signal measured by the sensor is the mixed signal of each vibration source, and contains strong noise at the same time. In this paper, a new separation method for mixed vibration signals in strong noise environment(SNR=-5) is proposed. Firstly, the time-delay auto-correlation de-noising method is used to de-noise the mixed signals, and then the common blind separation algorithm (MSNR algorithm is used here) is used to separate the mixed vibration signals, which improves the separation performance. The simulation results verify the validity of the method. The proposed method provides a new idea for health monitoring and fault diagnosis of engine rotor vibration signals.

## 1  Introduction

During the operation of rotating machinery, the changes of physical parameters such as vibration and noise will inevitably occur. These changes are often the early fault factors leading to engine failure. The vibration signal measured by the sensor installed on the rotating machinery is a mixture of several vibration signals. How to analyze, process and identify these signals is very important for judging the working state of rotating machinery and fault diagnosis. Various traditional modern signal processing methods, such as Fourier transform, short-time Fourier transform and wavelet transform, have been widely used in vibration signal analysis. However, for mixed vibration signals in rotating machinery, the above analysis methods have obvious shortcomings, and it is difficult to separate or extract source signals independently.

Blind source separation(BSS) technology can separate multiple mixed signals, and the separated output signal will not lose the weak feature information in the source signal. The seminal work on blind source separation is by Jutten and Herault[1] in 1985, the problem is to extract the underlying source signals from a set of mixtures, where the mixing matrix is unknown. In other words, BSS seeks to recover original source signals from their mixtures without any prior information on the sources or the parameters of the mixtures. Its research results have been widely applied in many fields, such as speech recognition, wireless communication, biomedicine, image processing, vibration signals separation, and so on.[2–5]

There have been many effective and distinctive blind source separation algorithms, including fast fixed-point algorithm, natural gradient algorithm, EASI algorithm and JADE algorithm. When separating noiseless mixed signals, these algorithms show good separation performance. However, when the signal-to-noise ratio of the noisy signal is very low, the separation performance will become very poor, because these algorithms are derived without considering the noise model. Noise is ubiquitous, its existence not only has a serious impact on the normal work of the system, but also affects the normal measurement of useful signals. In signal processing, in order to retain useful signals, people always try their best to remove background noise. So the research of signal detection, especially the extraction and detection of weak signals submerged in strong noise, is a common problem that many engineering applications face and need to solve urgently.

In the process of machine operation, the vibration signal measured by vibration sensor will inevitably contain noise signal. When the Blind Source Separation (BSS) algorithm is used to separate the mixed vibration signals directly, it may cause great errors or draw wrong conclusions. Therefore, noise reduction is particularly important before blind separation of mechanical vibration signals.

Many scholars have used the combination of wavelet de-noising and blind source separation to separate mixed signals in noisy environment, and achieved some results. However, the wavelet de-noising method needs to set threshold, which may remove weak signals of useful components in mixed signals, leading to wrong separation results. Time-delay auto-correlation de-noising method is widely used in the de-noising of rotor vibration signals, and it does not lose useful components in the de-noising process.

Nowadays, there have been lots of blind source separation algorithms to calculate a de-mixing matrix, so we can make the estimated source signal only by the received signal. In this paper we select and optimize the blind source separation algorithm based on MSNR.[7] It has very low computational complexity because de-mixing matrix can be achieved without any iterative.

In this paper, the time-delay auto-correlation method is used to de-noise the noisy mixed signal, and then the MSNR algorithm is used to separate the de-noised mixed signal. The separation effect is further improved.

The rest of the paper is organized as follows. In Section 2, we introduce the noisy signal BSS model and principle of the time-delay auto-correlation method, the improved MSNR algorithm is summarized in the end. In Section 3, the simulation experiment that indicates the effectiveness of the method is presented. The final section is a summary of the content of this paper and possible application areas.

## 2  Methodology

*2.1  Noisy Signal BSS Model*

Source signals $s_i(t)$ come from different signal sources (assumes that the signal is continuous signal), so $s_i(t)$ can be think mutual statistical independence, As shown in Fig.1, $x_i(t)$ is mixed signals or observation signals.

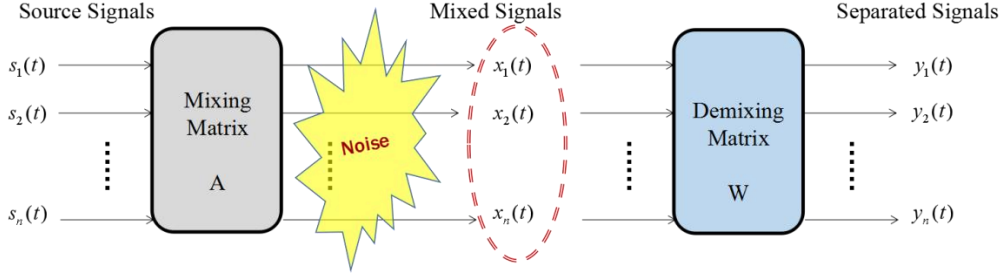

**Fig. 1** Noisy blind source separation model

The problem of basic linear BSS can be expressed algebraically as follows:

$$x_i(t) = \sum_1^n a_{ij}(s_i(t) + v(t)) \tag{1}$$

Where $a_{ij}$ is mixed coefficient, formula (1) can be write in vector as follow:

$$x(t) = A(s(t) + v(t)) \tag{2}$$

Where $s(t) = [s_1(t) \cdots s_n(t)]^T$ is a column vector of source signals, $x(t) = [x_1(t) \cdots x_n(t)]^T$ is vector of mixed signals or observation signals, $v(t)$ is additive white Gaussian noise, which is a basic noise model used in information theory to mimic the effect of many random processes that occur in nature. $A$ is $n \times n$ mixing matrix. Problem of BSS only know observation signals and statistical independence property of Source signals. In virtue of the knowledge of probability distribution of Source signals we can recover Source signals. Assume $W$ is $n \times n$ de-mixing matrix or separating matrix, problem of BSS can be describe as follow:

$$y(t) = Wx(t) \tag{3}$$

Where $y(t)$ is a estimate of or separated signals. BSS has two steps, firstly, create a cost function $F(W)$ with respect to $W$, if $W'$ can make $F(W)$ reach to maximum, $W'$ is the de-mixing matrix. Secondly, find a effective iterative algorithm for solution of $\partial F / \partial W = 0$. In this paper, cost function is the function of signal noise ratio, optimize processing of cost function result in generalized eigenvalue problem, de-mixing matrix was achieved by solving the generalized eigenvalue problem without any iterative.

*2.2 MSNR Algorithm*

Maximum signal-to-noise ratio (MSNR) algorithm belongs to matrix eigenvalue decomposition method [7]. By constructing the signal-to-noise ratio contrast function and estimating the separation matrix by eigenvalue decomposition or generalized eigenvalue decomposition, the closed-form solution can be found directly without iterative optimization process. Therefore, it has the advantages of simple algorithm and fast running speed, and is convenient for real-time processing and hardware implementation of FPGA. The time continuous radio signal is sampled and changed into a discrete value. In the following formula, the time mark $t$ becomes $n$.

According to the model of blind source separation, the error $e(n) = s(n) - y(n)$ between the source signal $s(n)$ and the output signal $y(n)$ is regarded as noise. When the minimum value of $e(n)$ is taken, the estimated value $y(n)$ is the optimal approximation of the source signal $s(n)$, and the effect of blind source separation is the best. The power ratio of source signal $s(n)$ to $e(n)$ is defined as signal-to-noise ratio. When $e(n)$ is the smallest, it is equivalent to the largest signal-to-noise ratio. According to this estimation criterion, the signal-to-noise ratio function[7] is constructed as follow:

$$F_{SNR} = 10\log\frac{s \cdot s^T}{e \cdot e^T} = 10\log\frac{s \cdot s^T}{(s-y)\cdot(s-y)^T} \tag{4}$$

Because the source signal $s(n)$ is unknown, the mean value of noise is 0, so we use moving average of estimate signals $\bar{y}(n)$ instead of source signals $s(n)$. Formula (4) can be write as:

$$F_{SNR} = 10\log\frac{s \cdot s^T}{e \cdot e^T} = 10\log\frac{\bar{y} \cdot \bar{y}^T}{(\bar{y}-y)\cdot(\bar{y}-y)^T} \tag{5}$$

Where $\bar{y}_i(n)$ is moving average of estimate signals $y(n)$. We replace $\bar{y}(n)$ with $y(n)$ in the molecule of formula (5) to simplify calculation, so we gained maximum signal noise ratio cost function as follow:

$$F_{SNR}* = 10\log\frac{y \cdot y^T}{(\bar{y}-y)\cdot(\bar{y}-y)^T} \tag{6}$$

According to formula (3), we get the formula (7) as follows.

$$\bar{y}(n) = W\bar{x}(n) \tag{7}$$

Where $\bar{x}(n)$ is a moving average of mixed signals $x(n)$. The definition uses the moving average algorithm to predict the source signal. We substitute formula (3) and formula (7) into formula (6) and Formula (8) is deduced.

$$F_{SNR}*(W) = 10\log\frac{y \cdot y^T}{(\bar{y}-y)\cdot(\bar{y}-y)^T} = 10\log\frac{Wx \cdot x^T W^T}{W(\bar{x}-x)\cdot(\bar{x}-x)^T W^T}$$
$$= 10\log\frac{WCW^T}{W\bar{C}W^T} = 10\log\frac{V}{U} \tag{8}$$

Where $\bar{C} = (\bar{x}-x)(\bar{x}-x)^T$ and $C = xx^T$ are correlation matrixs, $U = W\bar{C}W^T$, $V = WCW^T$.

## 2.3 Derivation of Separation Algorithms

According to formula (8), derivative of $F_{SNR}*(W)$ with respect to is:

$$\frac{\partial F_{SNR}*(W)}{\partial W} = \frac{2W}{V}C - \frac{2W}{U}\overline{C} \quad (9)$$

According to the definition, when the maximum value of the function $F_{SNR}*(W)$ is obtained, the gradient is 0. So we get the following formula.

$$WC = \frac{V}{U}W\overline{C} \quad (10)$$

We can obtain de-mixing matrix $W'$ by solving formula (10), it has been proved solution of formula (10) that is eigenvector of $\overline{C} \cdot C^{-1}$.[8] All source signals can be recovered once: $y = W'x$, where each row of $y$ corresponds to exactly one extracted signal $y_i$.

## 2.4 Auto-correlation De-noising

The auto-correlation function describes the relationship of the same signal at different times. For signal $x(t)$, its auto-correlation function is defined as:

$$R_x(\tau) = \lim_{T \to \infty} \frac{1}{T} \int_0^T x(t)x(t+\tau)dt \quad (11)$$

Where $\tau$ is the time delay of auto-correlation function, $T$ is the period of the signal. Formula (11) shows that the auto-correlation function of the periodic signal is the same period as that of the original signal. However, noise signals are generally uncorrelated. When the time delay is zero, the maximum auto-correlation value is obtained and tends to zero with the increase of the time delay. Therefore, the auto-correlation function can be used in the noise reduction of mechanical vibration signal, so as to retain the useful periodic signal in the vibration signal, effectively remove the random aperiodic white Gaussian noise, and achieve remarkable noise reduction effect.

The auto-correlation function values of white Gaussian noise and rotor vibration signals are shown in Fig. 2. When the vibration periodic signal contains Gauss white noise, the auto-correlation value is the largest near this condition $\tau = 0$, which is affected by noise. Therefore, we can remove some auto-correlation data near the condition $\tau = 0$ during removing noises.

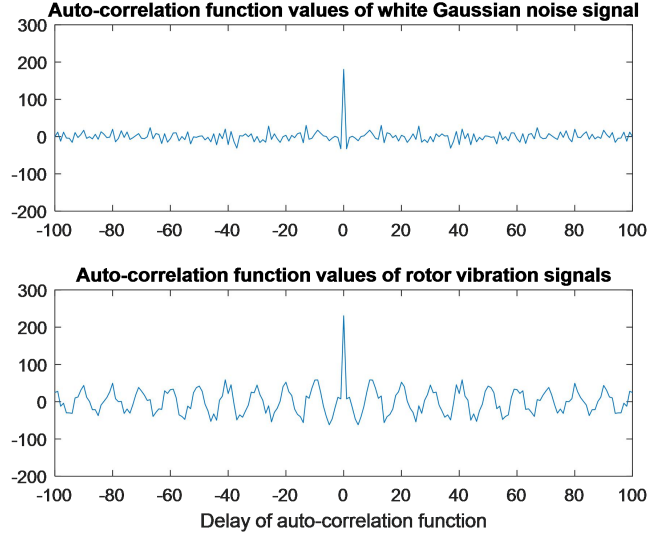

**Fig. 2** Auto-correlation function values of white Gaussian noise signal and rotor vibration signals

The improved MSNR algorithm based on auto-correlation de-noising can be summarized as: (1) Finding the auto-correlation function of noisy mixed signals $x(t)$. (2) Removing the data near the condition $\tau = 0$ and using the remaining data $\hat{x}(t)$ as the data of blind separation. (3) Blind separation of de-noised mixed signals $\hat{x}(t)$ by MSNR algorithm.

## 3 Experiments and Results

In order to verify the effectiveness of the algorithm, two sinusoidal periodic signals with different frequencies are used to simulate the mixing of vibration signals caused by different rotors. After the original vibration signal $s(t)$ is superimposed with Gaussian white noise whose signal-to-noise ratio is -5dB, the source signal completely submerged by a strong noise is more difficult to be restored and identified in the engineering fields.[9] The noisy mixed signal $x(t)$ is obtained by random mixing matrix $A$ (such as $A$=[0.4684 0.1952; 0.7384 0.5483]) . The number of samples N=1000.

Evaluating the performance of blind source separation, a correlation coefficient $C$ is introduced as a performance index.[10]

$$C(x,y) = \frac{\text{cov}(x,y)}{\sqrt{\text{cov}(x,x)}\sqrt{\text{cov}(y,y)}} \qquad (12)$$

$C(x,y) = 0$ means that x and y are uncorrelated, and the signals correlation increases as $C(x,y)$ approaches unity, the signals become fully correlated as $C(x,y)$ becomes unity.

Here are two experiments. In the first experiment, the noisy mixed signals $x(t)$ are separated directly by the MSNR algorithm, the separation results are shown in Fig. 3. After separation, the correlation coefficients between the separated signals and the sources are 0.4978 and 0.4806 respectively, the separation effect is not good and it is very difficult to recognize separated signals correctly.

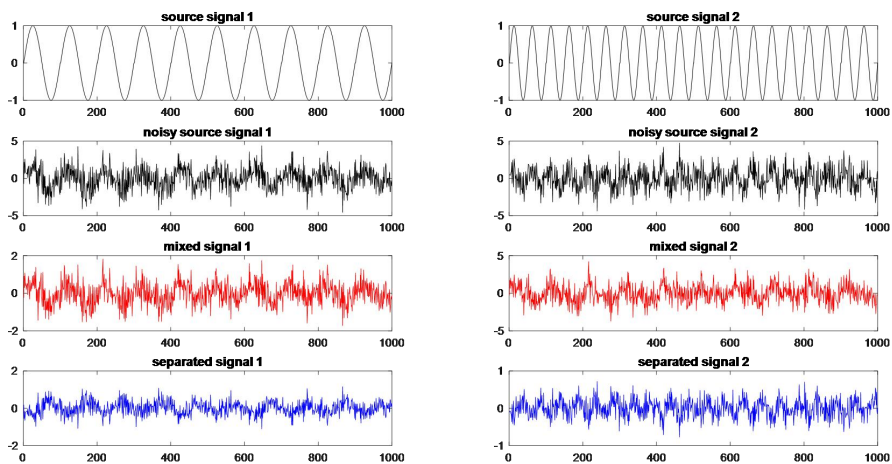

**Fig. 3**　Separation of noisy mixed signals by the MSNR algorithm(SNR=-5dB)

In the second experiment, the noisy mixed signals $x(t)$ are separated by the improved MSNR algorithm, the separation results are shown in Fig. 4. After separation, the correlation coefficients between the separated signals and the sources are 0.9987 and 0.9988 respectively, the sources are well recovered and the separation effect has been significantly improved.

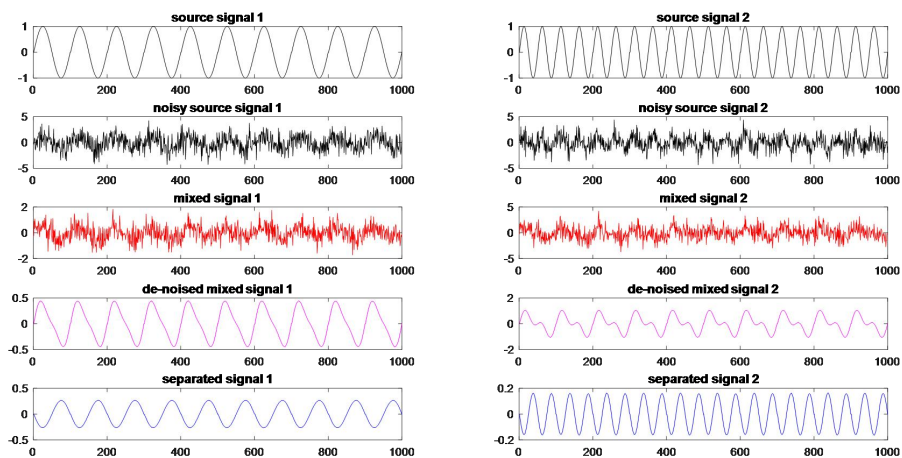

**Fig. 4**　Separation of de-noised mixed signals by the improved MSNR algorithm(SNR=-5dB)

By comparing the two experiments, it is fully demonstrated that time-delay correlation de-noising can effectively remove noise and improve signal-to-noise ratio, which provides the precondition for the accurate realization of blind source separation of noisy mixed signals.

## 4 Conclusions

Aiming at blind source separation of rotor vibration signals in high-noise environments, an improved MSNR algorithm is proposed in this paper. Blind separation of mixed signals with strong noise can lead to large errors or even incorrect separation results. The time-delay auto-correlation de-noising method can effectively remove the strong noise signal without losing the useful components of the original signal, which greatly improves the signal-to-noise ratio and provides the precondition for the accurate realization of blind separation. It provides a new method for separating mixed signals in strong noise environment and further expands the applicability of the MSNR algorithm. Due to its simple principle and good transplantation capability, it can be applied to the vibration signals of various mechanical rotors, such as the separation and detection of vibration signals of aero-engine and internal combustion engine.